\documentclass[alpha-refs]{wiley-article}

\usepackage{siunitx}
\usepackage{blindtext}
\usepackage{hyperref}
\papertype{Original Article - Draft}
\paperfield{}

\title{Salinity-induced limits to Mangrove canopy height and diversity}

\author[1\authfn{1}]{Saverio Perri}
\author[1,2]{Matteo Detto}
\author[1,3]{Amilcare Porporato}
\author[4\authfn{1}]{Annalisa Molini}

\contrib[\authfn{1}]{Corresponding authors.}

\affil[1]{High Meadows Environmental Institute, Princeton University, Princeton, NJ, 08544, USA}
\affil[2]{Ecology \& Evolutionary Biology, Princeton University, Princeton, NJ, 08540, USA}

\affil[3]{Civil \& Environmental Engineering, Princeton University, Princeton, NJ, 08540, USA}

\affil[4]{River-Coastal Science \& Engineering, Tulane University, New Orleans, LA, 70118, USA}

\corraddress{Saverio Perri, High Meadows Environmental Institute, Princeton University, Princeton, NJ, 08544, USA \\
Annalisa Molini, River-Coastal Science \& Engineering, Tulane University, New Orleans, LA, 70118, USA}
\corremail{sperri@princeton.edu, amolini@tulane.edu}

\fundinginfo{This study was supported by the US Department of Energy Terrestrial Ecosystem Science Program under grant no. DE-SC0020116. S.P. and A.P. also acknowledge the support provided by the Princeton University’s Dean for Research, High Meadows Environmental Institute, Andlinger Center for Energy and the Environment, and the Office of the Provost International Fund.}

\runningauthor{Perri et al.}

\begin{document}

\begin{frontmatter}
\maketitle

\begin{abstract}
Mangrove canopy height is a key metric to assess tidal forests' resilience in the face of climate change.
In terrestrial forests, tree height is primarily determined by water availability, plant hydraulic design, and disturbance regime.
However, the role of water stress remains elusive in tidal environments, where saturated soils are prevalent, and salinity can substantially affect the soil water potential.
Here, we use global observations of maximum canopy height, species richness, air temperature, and seawater salinity -- a proxy of soil water salt concentration -- to explain the causal link between salinity and Mangrove stature.\\
Our findings suggest that salt stress affects Mangrove height both directly, by reducing primary productivity and increasing the risk of xylem cavitation, and indirectly favoring more salt-tolerant species, narrowing the spectrum of viable traits, and reducing biodiversity.
Yet, salt tolerance comes to a price. Highly salt-tolerant mangroves are less productive and generally shorter than more sensitive species, suggesting a causal nexus between salinity, biodiversity, and tree height.
As sea-level rise enhances coastal salinization, failure to account for these effects can lead to incorrect estimates of future carbon stocks in Tropical coastal ecosystems and endanger preservation efforts.

% Please include a maximum of seven keywords
\end{abstract}
\end{frontmatter}

% \linenumbers*[1] 
%\linenumbers
\section{Introduction}
\begin{figure*}[t]%b%h
\centering 
\includegraphics[width=1\textwidth]{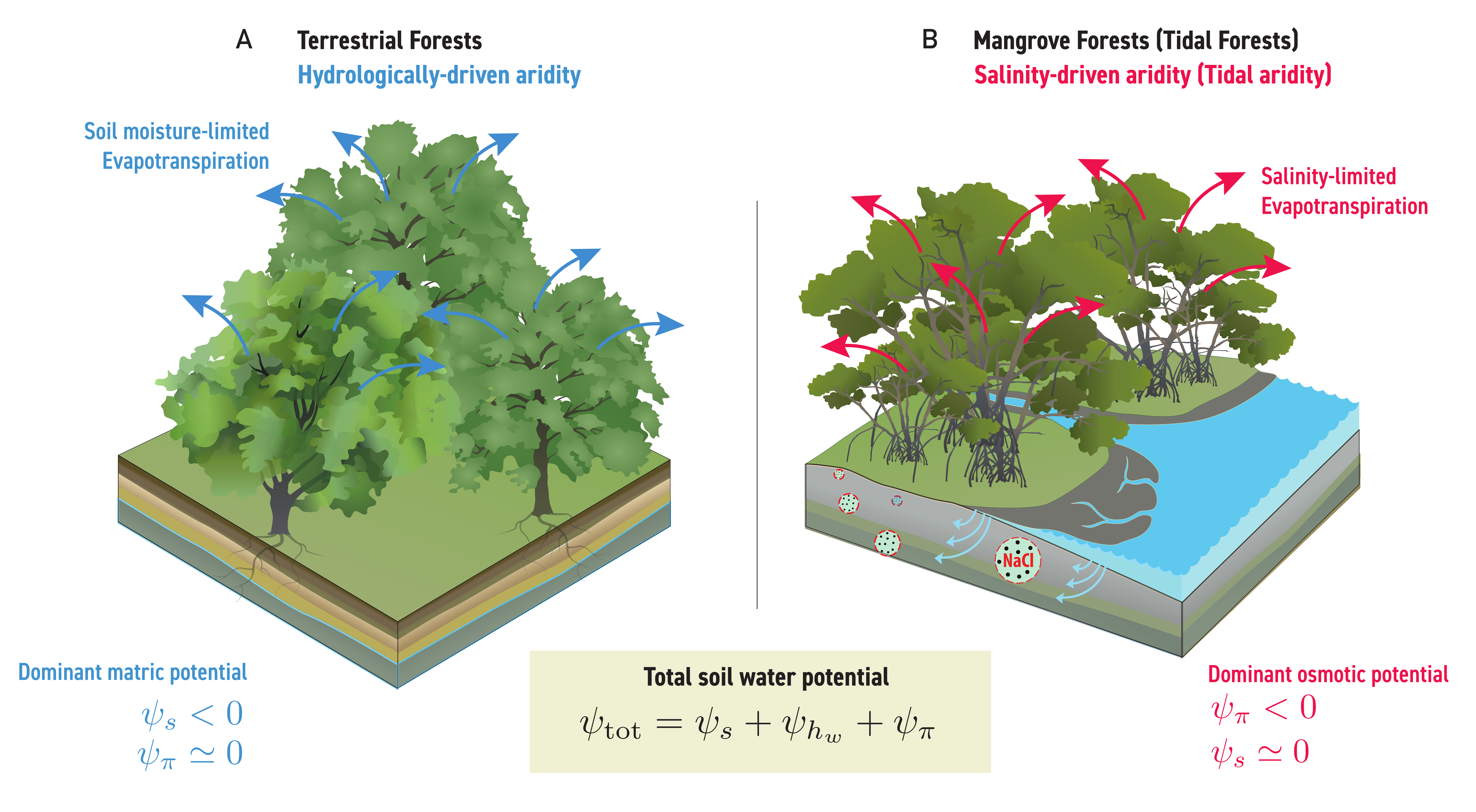}
\caption{{\textbf{Hydrologically-driven versus salinity-driven hydraulic stress.}} Conceptual representation of water stress in terrestrial (A) and tidal (B) forests, as regulated by the total soil water potential $\psi_{\rm{tot}}$, defined as the sum of matric potential $\psi_s$, osmotic potential $\psi_{\pi}$, and gravimetric potential $\psi_{h_w}$. In terrestrial forests (A), water movements in the soil are mainly controlled by $\psi_s$ and soil moisture patterns (\textit{hydrologically-driven hydraulic stress}). In contrast, in tidal environments (B), soils are primarily saturated ($\psi_{s} \simeq 0$), and the contribution of $\psi_{\pi}$ becomes dominant (\textit{salinity-driven hydraulic stress}).}
  \label{fig:Figure1}
\end{figure*}

%%==================================%%
%%          INTRODUCTION            %%
%%==================================%%
%\section{Introduction}\label{sec1} 
Canopy height is a critical variable in forest ecology. It regulates access to light, promotes diversity in plant functional types, and enhances ecosystem resilience to environmental disturbances
~\citep{Walker1999,Falster2003,Giardina2018,detto2021maintenance}.
When interpreted through allometric relations, the height of trees also represents an essential parameter to estimate aboveground carbon stocks and design interventions to protect critically endangered ecosystems~\citep{Saenger1993,Moles2009,worthington2020harnessing}.
In terrestrial forests, the height of trees is primarily controlled by water availability and biotic and abiotic factors, such as light competition, micro-climate, species allocation strategies, and the disturbance regime~\citep{Koch2004,Moles2009,Dybzinski2011,Klein2015,Tao2016,Giardina2018}.\\
\indent In contrast, in Mangrove forests thriving at the interface between the terrestrial and the marine environment, tidal inundation and river discharge maintain nearly saturated soil conditions, and physical water availability is far from being a limiting factor~\citep{Ball1988,Field1998}. 
While alternative forms of water stress, like the limiting effects of pore water salinity on root water uptake and transpiration, could play a primary role in regulating Mangrove height, the broad mechanisms linking tidal drivers to canopy height remain largely unexplored.\\
\indent Maximum canopy height, $H_{\rm{max}}$, a parameter relatively easy to obtain from forest inventories, airborne LiDAR, or other remote sensing platforms~\citep{Lucas2002,Heumann2011,Fatoyinbo2013,Salum2020}, has been extensively used to investigate the relation between Mangrove height and environmental drivers at the global scale~\citep{Friess2019,Simard2019,charrua2020assessment}.
These studies have primarily focused on understanding how the latitudinal patterns of $H_{\rm{max}}$ are affected by both fundamental climatic factors like precipitation and temperature and sporadic ecological disturbances such as tropical storms~\citep{feher2017linear,Simard2019,charrua2020assessment}.
Still, major tidal stress factors as soil pore water salinity display a large regional variability~\citep{rovai2021macroecological} and can hardly be typified through zonal averages. 
Therefore, their effects remain more challenging to pinpoint in global studies.\\
\indent Salinity has been long-known as one of the dominant sources of abiotic stress in coastal wetlands, regulating local carbon allocation and production~\citep{Ball1988,alongi1998coastal,Perri2019}.
Gradients of salinity have been linked to tidal zonation, species succession and abundance at the ecosystem scale~\citep{GreinerLaPeyre2001,Wendelberger2017,Perri:2018ur}.
Moderate salinity is generally associated with high species richness  \citep{Ball1998richness,islam2016species}, which has been related to elevated productivity \citep{mittelbach2001observed,bai2021mangrove}. While the relationship between species diversity and productivity remains controversial and may depend on the spatial scale of interest \citep{whittaker2003observed,mensah2018diversity}, biotic interactions (e.g., competition for light or ecological facilitation) promote tall canopies in highly diverse Mangrove forests \citep{wolf2012plant,meyer2020canopy}. Affecting species richness, salinity may, therefore, exert an indirect control on canopy height.\\ 
% \textbf{The highest species richness across Mangrove ecosystems has been observed under moderate salinity \citep{Ball1998richness,islam2016species}. High species richness, in turn, is generally related to high productivity \citep{mittelbach2001observed,bai2021mangrove}, even though this relationship remains controversial and may depend on the spatial scale of interest \citep{whittaker2003observed,mensah2018diversity}. \\} 
\indent At the same time, salinity is a primary hydrological driver, known to reduce soil water potential, limit root water uptake, increase the risk of xylem cavitation, and impair photosynthesis~\citep{munns2002comparative,perri2018plant,Perri2019,Perri:2022NatGeo}.
Recent studies on Mangrove biomass stocks and allometric relations have suggested that aridity and tidal inundation frequency -- which are both strongly correlated with water salinity -- are dominant limiting factors for coastal ecosystems' carbon storage~\citep{adame2020Mangroves,Bathmann2021,rovai2021macroecological}.
These effects are analogous to those of water stress in terrestrial forests, and similarly to water stress, they could exert significant controls on species competition and canopy height~\citep{Koch2004,Klein2015}. While in terrestrial forests the soil water potential, $\psi_{\rm{tot}}$, is governed by soil moisture availability ({\it{hydrologically-driven hydraulic stress}}; Figure~\ref{fig:Figure1}A), in saturated tidal soils the matric potential, $\psi_{s}$, is close to zero, and osmotic effects largely regulate access to water ({\it{salinity-driven hydraulic stress}}; Figure~\ref{fig:Figure1}B).
Accordingly, salinity should be regarded as a form of {\it{tidal aridity}}, capable of imposing major constraints to Mangrove canopy dynamics and productivity.\\
\indent Here we investigate the relation between pore water salinity and Mangrove height across scales ranging from regional to global. 
Our main hypothesis is that the interplay of salt stress and plant salt tolerance can be used to explain and generalize this relation across a large variety of coastal ecosystems worldwide -- in analogy with the effects of water stress in terrestrial ecosystems. 
We also hint at the causal relation between salt-stress, species succession, and species richness as one of the primary mechanisms through which salinity can affect Mangrove height at regional and global scales.
%%%%========================END INTRODUCTION
%%%%========================BEGIN METHODS
\section{Data and Methods}\label{sec:D&M}
\subsection{Data Sources}
  \label{sec:data}  
Our analysis relies on the global Mangrove maximum canopy height ($H_{\rm{max}}$) dataset generated by~\citet{Simard2019global} by integrating the Shuttle Radar Topography Mission (SRTM) digital elevation model \citep{Farr2007} with the LiDAR heights from the Geoscience Laser Altimeter System (GLAS) mission \citep{Fatoyinbo2008}. The dataset, which also provides global estimates of Mangrove distribution and aboveground biomass, is made publicly available from the Oak Ridge National Data Archive (ORNL DAAC) at a native 30-meter spatial resolution. Both SRTM and GLAS retrieval refers to the year 2000.
To limit Mangroves’ misattribution away from coastal areas, we masked the raw Mangrove canopy height data with the spatial distribution of Mangroves reported in the World Atlas of Mangroves \citep{spalding2010world}. \\
\indent The World Atlas of Mangroves was also used to infer global patterns of species richness, $SR$, by overlaying the spatial coverage at 30 meters resolution of the 60 Mangroves species listed in the Atlas. 
$SR$ is estimated as the number of species with the same bio-geographical distribution, and represents a potential $SR$. The actual number of species per plot is expected to be lower, as fast-growing (and potentially taller) species out-compete the less productive ones.\\ 
\indent Maximum canopy height was investigated as a function of historical mean air temperature ($T_{\rm{air}}$; representing here climatic forcing) and sea-surface salinity ($S_{\rm{sw}}$). While other environmental variables such as precipitation and evapotranspiration are expected to be correlated with pore water salinity \citep{kelble2007salinity,rovai2021macroecological}, the salt mass balance of a Mangrove ecosystem depends not only on these local environmental conditions but also on large-scale ocean circulation and the quantity and quality of riverine inputs \citep{lane2007effects,gomez2019enso}. Given such complexity and the lack of direct measurements of soil salinity, $S_{\rm{sw}}$ represents the best available proxy for salinity in coastal ecosystems and was, therefore, selected for the analysis performed here.
Air temperature, $T_{\rm{air}}$ across the Mangrove biogeographical range (39$^{\circ}$ S -- 30$^{\circ}$ N) was obtained from the WorldClim dataset \citep{Fick2017} over the period 1970--2000, while sea-surface salinity, $S_{\rm{sw}}$ --  was derived from the GLORYS12V1 reanalysis~\citep{GLORYS12V1} for the years 1993-2018.
WorldClim and GLORYS12V1 products are provided as monthly averages at a 30 arc-seconds and 300 arc-seconds spatial native spatial resolution, respectively.
Although $S_{\rm{sw}}$ and $T_{\rm{air}}$ time spams are not completely overlapping, their temporal extent is sufficient to derive long-term means that can be used as explanatory variables for $H_{\rm{max}}$. 
All the data were aggregated at 10-minutes spatial resolution (600 arc-seconds, about 18.5 km at the equator), and average values were obtained for each pixel.\\
\indent Mangrove canopy height was associated with mean salinity and air temperature based on a proximity analysis. This was obtained through the use of a focal statistic where, for each $H_{\rm{max}}$ pixel, the corresponding $S_{\rm{sw}}$ and $T_{\rm{air}}$ were calculated as average values among neighboring cells (up to 5$\times$5 cells).
%%%%%%%%%%%%%%%
%%
%%%%%%%%%%%%%%%
\subsection{Latitude-dependent and Site-dependent analyses}
  \label{sec:LD and SD}
The relation between mangrove canopy height, salinity, air temperature  and species richness was explored both zonally (\textit{latitude-dependent analysis}) and locally (\textit{site-dependent analysis}).
In the latitude-dependent analysis, each variable was then averaged zonally over 1$^{\circ}$ of latitude, reducing the sample size to 70 averages corresponding to the observed geographical range of Mangrove ecosystems (39$^{\rm{o}}$ S -- 30$^{\rm{o}}$ N). The zonal averages where calculated at the global scale and across the different bio-geographic regions: (a) Americas and West Africa (AWA: 120$^{\rm{o}}$ W - 13$^{\rm{o}}$ E), (b) East Africa and Middle East (EAME: 30$^{\rm{o}}$ E - 77$^{\rm{o}}$ E), and (c) Indo-Pacific Asia (IPA: 78$^{\rm{o}}$ E - 152$^{\rm{o}}$ W; Figure \ref{fig:Figure1}E-G). 
The site-dependent analysis, in contrast, was performed comparing the local gridded $H_{\rm{max}}$, $S_{\rm{sw}}$, and $T_{\rm{air}}$ data to $SR$ estimates. 
We explored how the main statistical descriptors (i.e., median, 25$^{\rm{th}}$ and 75$^{\rm{th}}$ percentiles, min and max non-outlier values) of $H_{\rm{max}}$, $S_{\rm{sw}}$, and $T_{\rm{air}}$ vary as a function of $SR$.\\
Finally, to reduce the variability of $H_{\rm{max}}$ within the dataset and partially decouple the direct impact of salinity on canopy height from the indirect effect mediated by species richness, the relation between canopy height and environmental variables was investigated analyzing the mean values of $H_{\rm{max}}$ as a function of mean $S_{\rm{sw}}$, and $T_{\rm{air}}$, conditional to species richness ($\langle H_{\rm{max}} \mid SR \rangle$). 
The $\langle H_{\rm{max}} \mid SR \rangle$ was also related to the $S_{\rm{sw}}$ coefficient of variation, $CV_{S_{\rm{sw}}}$, and a modified $CV$ for air temperature, $CV^*_{T_{\rm{air}}}$. 
The latter was calculated with respect to the optimal temperature value $T_{\rm{opt}}$, identified around the mode of the global $T_{\rm{air}}$ distribution in Mangrove ecosystems ($\sim$27$^\circ$C; see Section \ref{sec:Res_bio}). 
We first computed the mean deviation ($\sigma^*_T$) around $T_{\rm{opt}}$. 
The modified coefficient of variation was then estimated as $CV^*_{T_{\rm{air}}}$=$\sigma^*_T / T_{\rm{opt}}$, which reflects the seasonal variability of the mean air temperature around its optimal value.
%\indent Although our analysis demonstrates that $S_{\rm{sw}}$ is a good predictor of $H_{\rm{max}}$ at the local scale, canopy height is expected to be even more correlated to soil pore water salinity, which reflects the local-scale hydrological regime and geomorphology \cite{Ward2006,Krauss2008}. 
%However, soil salinity in coastal ecosystems remains challenging to estimate through remote sensing measurements, and a global appraisal is still missing. For this reason, the relative contribution of salinity has been quantified through $S_{\rm{sw}}$ that, over long  temporal scales, can be considered a reasonable proxy for soil salinity in coastal ecosystems.
%========================================%
% Analysis of the variance
%========================================%
\subsection{Analysis of variance}
\label{sec:Variance}
Previous studies have quantified the impact of $T_{\rm{air}}$, $S_{\rm{sw}}$ and precipitation on Mangroves biomass stocks by performing ordinary least squares regression (OLSR) analyses \citep{Rovai2018,Ribeiro2019,Simard2019}. However, OLSRs can yield ambiguous results in case of high multicollinearity \citep{Naes1985} that is expected, for example, between $T_{\rm{air}}$ and $SR$. 
Here, the significance of the relations between $H_{\rm{max}}$ and the selected environmental drivers ($S_{\rm{sw}}$, $T_{\rm{air}}$, and $SR$) was tested through multivariate regression analysis. 
We opted for a partial least-squares regression (PLSR) model, which has the advantage of accounting for the possible multicollinearity among the predictor variables \citep{Rosipal2005}. \\
\indent The selection of environmental variables to include in the PLSR model was based on the results of previous studies \citep{Simard2019,rovai2021macroecological} and on the observation that rainfall, tidal regime, and evapotranspiration, rather than directly affecting mangroves productivity, are all proxies of salinity \citep{fosberg1961vegetation,Field1995}. \citet{Simard2019}, in particular, considered a wide range of bioclimatic variables and concluded that the latitudinal variation of maximum canopy height can be explained by just average temperature and average annual precipitation. Species richness was not included in their regression model. \citet{rovai2021macroecological} suggested that Mangrove biomass stocks are controlled by local or regional forcings rather than latitude-dependent averages. They found that aridity, tidal amplitude, and duration, all proxies of salinity, explain most of the $H_{\rm{max}}$ variance when analyzed at the ecosystem (local or pixel) scale. Based on these previous findings, and to test our hypothesis that both salinity species richness could affect Mangrove canopy height, we selected $S_{\rm{sw}}$, $T_{\rm{air}}$, and $SR$ as plausible explanatory variables for both the latitude-dependent and site-dependent analyses.\\
\indent The PLSR, performed running the SIMPLS algorithm \citep{de1993simpls}, was used to explain the observed variance in both the latitude- and site-dependent patterns. 
We first used a univariate regression model to estimate the variance explained by each variable alone. We then combined the three variables through the multivariate PLSR.
The best PLSR models (i.e., the ones that can explain most of the variance) were obtained by including all the variables in the latitudinal analysis. We excluded  $T_{\rm{air}}$ from the site-specific model because of its high correlation with $SR$ (see Section \ref{sec:Forcings}).
%The regression model results in the latitudinal analysis show that $SR$ alone explains up to 80$\%$ of the variance. $S_{\rm{sw}}$ and $T_{\rm{air}}$ alone can explain up to 52$\%$ and 35$\%$ of the variability, respectively. However, when all the variables are considered together in the PLSR, $SR$ is shown to be the main driver for $H_{\rm{max}}$, and $S_{\rm{sw}}$ and $T_{\rm{air}}$ are only weakly related to canopy height.\\
%At the site-specific scale, $SR$ remains the best predictor explaining up to 34$\%$ of the variance. Yet, $S_{\rm{sw}}$ is shown to be a significant variable, explaining up to 26$\%$ of the variance. When combined with $SR$ in the multivariate model, the significance of $S_{\rm{sw}}$ as an explanatory variable remains, while $T_{\rm{air}}$ is only indirectly related to $H_{\rm{max}}$ through its control on $SR$.
%%%%========================END METHODS
%%%%========================
%%%%========================BEGIN METHODS
%%==================================%%
%%              RESULTS             %%
%%==================================%%
\section{Results and discussion}\label{results}
%================================================================%
% Global and regional analysis
%================================================================%
\subsection{Regional versus global salinity controls on Mangrove canopy height}
In line with previous studies performed in terrestrial forests, many authors have focused on the relation between Mangrove height -- as a proxy of above-ground biomass -- and major climatic drivers, such as precipitation and air temperature~\citep{pickens2011temperature,Friess2019,Simard2019,charrua2020assessment}.
These studies consistently show that the latitudinal variability of Mangrove height strongly correlates with air temperature and precipitation, concluding that these climatic factors alone can explain most of the observed spatial variability of $H_{\rm{max}}$.
%%===============================================%
%%   Figure 2 - Global versus regional drivers 
%================================================%
\begin{figure*}[t!]
\centering 
\includegraphics[width=1\textwidth]{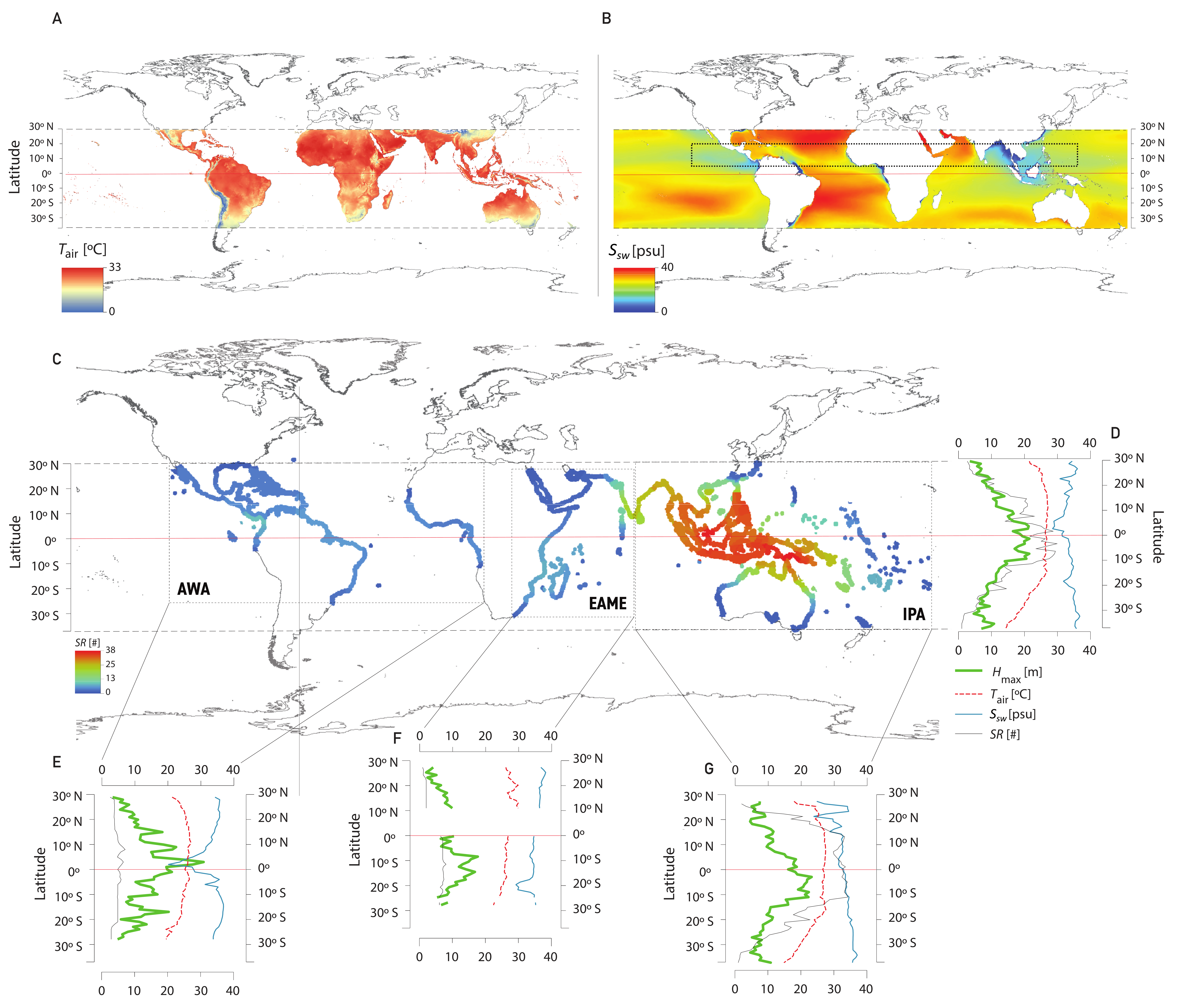}
%   \caption{\textbf{Global versus regional drivers of Mangrove canopy height}}
%\end{figure}
%\addtocounter{figure}{-1}
%\begin{figure} [t!]
\caption{\textbf{Global versus regional drivers of Mangrove canopy height:} Global multi-year average of A) mean air temperature ($T_{\rm{air}}$) from the WorldClim dataset \citep{Fick2017} for the period 1970-2000, and B) surface seawater salinity ($S_{\rm{sw}}$) from the GLORYS12V1 reanalysis over the period 1993-2018 \citep{GLORYS12V1}. % {\bf{[We should provide some sort of justification for the non overlapping range of years, check methods]}}.
 Data are shown for the latitude range 39$^{\rm{o}}$ S -- 30$^{\rm{o}}$ N, which spans the entire Mangrove biogeographical range as reported in the World Mangrove Atlas~\citep{spalding2010world}. The dashed rectangle in B) highlights areas of high regional salinity variability.  
 %Salinity patterns display  a less marked latitudinal dependence due to land distribution, the complexity of the global thermohaline circulation and riverine inputs.
 C) Global distribution Mangroves species richness ($SR$) derived from the World Mangrove Atlas~\citep{spalding2010world}. D-G) Latitudinal distribution of $H_{\rm{max}}$ (green lines), $T_{\rm{air}}$ (red lines), $S_{\rm{sw}}$ (light blue lines), and $SR$ (black lines) D) at a global scale, E) in the AWA, F) EAME, and G) IPA regions. The blue shaded area in F) indicates a range of EAME latitudes where data are too scarce to support the analysis.}
  \label{fig:Figure2}
\end{figure*}
%===================================END FIGURE 2
\indent However, the analysis of global correlations is only one part of the tale.
Rather than directly affecting canopy height, air temperature and precipitation might shape the latitudinal patterns of $H_{\rm{max}}$  indirectly through their control on aridity and species richness. 
There is a broad scientific consensus around the idea that the ecological role of air temperature is to determine the geographical range of Mangrove ecosystems, thus affecting biodiversity rather than directly influencing canopy height~\citep{clough1982physiological,Duke1998,Osland2016}.
Besides, temperature governs evapotranspiration, which with precipitation regulates the concentration of salts in the soil pore water and, more broadly, aridity~\citep{Osland2016,rivera2017advancing}.\\
\indent In the same way, it is unlikely that freshwater inputs from precipitation can significantly influence soil water availability and plant status in tidal areas, where soils are predominantly saturated~\citep{Ball1988,Field1998}. 
Although precipitation is related to river discharge and nutrient delivery~\citep{Twilley1999,Reef2010}, its primary role in coastal wetlands is to dilute soil water salinity~\citep{fosberg1961vegetation,Rodriguez2016}. 
As a result, it should be regarded as a proxy for salinity rather than a climatic factor affecting $H_{\rm{max}}$ directly~\citep{fosberg1961vegetation,Field1995,Gilman2008}.\\
\indent At the same time, salinity controls on tidal ecosystems are local in nature, making it challenging to establish a clear nexus between salt stress and canopy height at the global scale~\citep{Ward2006}.
Contrary to climatic drivers like average air temperature, which displays distinct latitudinal patterns in response to irradiance distribution 
(Figure~\ref{fig:Figure2}A), coastal seawater salinity can be affected by the local hydrological regime, coastal geomorphology, tidal controls and thermohaline circulation~\cite{Schmidt2004,Schmitt2008,Herbert2015} (Figure \ref{fig:Figure2}B). 
In addition, while mangroves are halophytes, their adaptations to salinity vary widely across the different species, resulting in a broad spectrum of physiological responses to salt stress \citep{parida2010salt,Reef2015regulation}.\\
\indent To pinpoint these local effects, we disaggregated global patterns of maximum Mangrove height ($H_{\rm{max}}$), sea surface salinity ($S_{\rm{sw}}$; a proxy for pore water salinity in coastal ecosystems), mean air temperature ($T_{\rm{air}}$; representing climatic forcing), and species richness ($SR)$ to the regional scale (see Methods). %(Figure~\ref{fig:Figure2}C-G).
Although consistent with previously observed latitudinal patterns, our latitude-dependent analysis suggests that $SR$ is the primary covariate of the observed $H_{\rm{max}}$ distribution, with $S_{\rm{sw}}$ and $T_{\rm{air}}$ only weakly correlated to canopy height (Figure \ref{fig:Figure2}D-G).
Due to its inhomogeneous latitudinal distribution, salinity appears to be a poor descriptor of $H_{\rm{max}}$ when analyzed at the global scale (Figure \ref{fig:Figure2}D).
However, it shows different -- and at times marked -- regional impacts on the AWA, EAME, and IPA Mangroves (Figure \ref{fig:Figure2}C-E).
In the AWA region (Figure \ref{fig:Figure2}E), characterized by a wide range of salinity conditions and intermediate biodiversity, $S_{\rm{sw}}$ appears to exert a clear influence on $H_{\rm{max}}$. 
In contrast, this relation is more elusive in the EAME region (Figure \ref{fig:Figure2}F) -- characterized by extreme salinity conditions and meagre to intermediate biodiversity -- and in the IPA Mangroves (Figure \ref{fig:Figure2}G), where low to moderate salinity conditions and high biodiversity represent the norm. 
These results might suggest that the influence of $S_{\rm{sw}}$ on $H_{\rm{max}}$ is, to some extent, mediated by species richness. 
%===================================================%
% Biodiversity control
%===================================================%
\subsection{Biodiversity-mediated salinity controls}
\label{sec:Res_bio}
\begin{figure*}[t!]
\centering 
\includegraphics[width=1.0\linewidth]{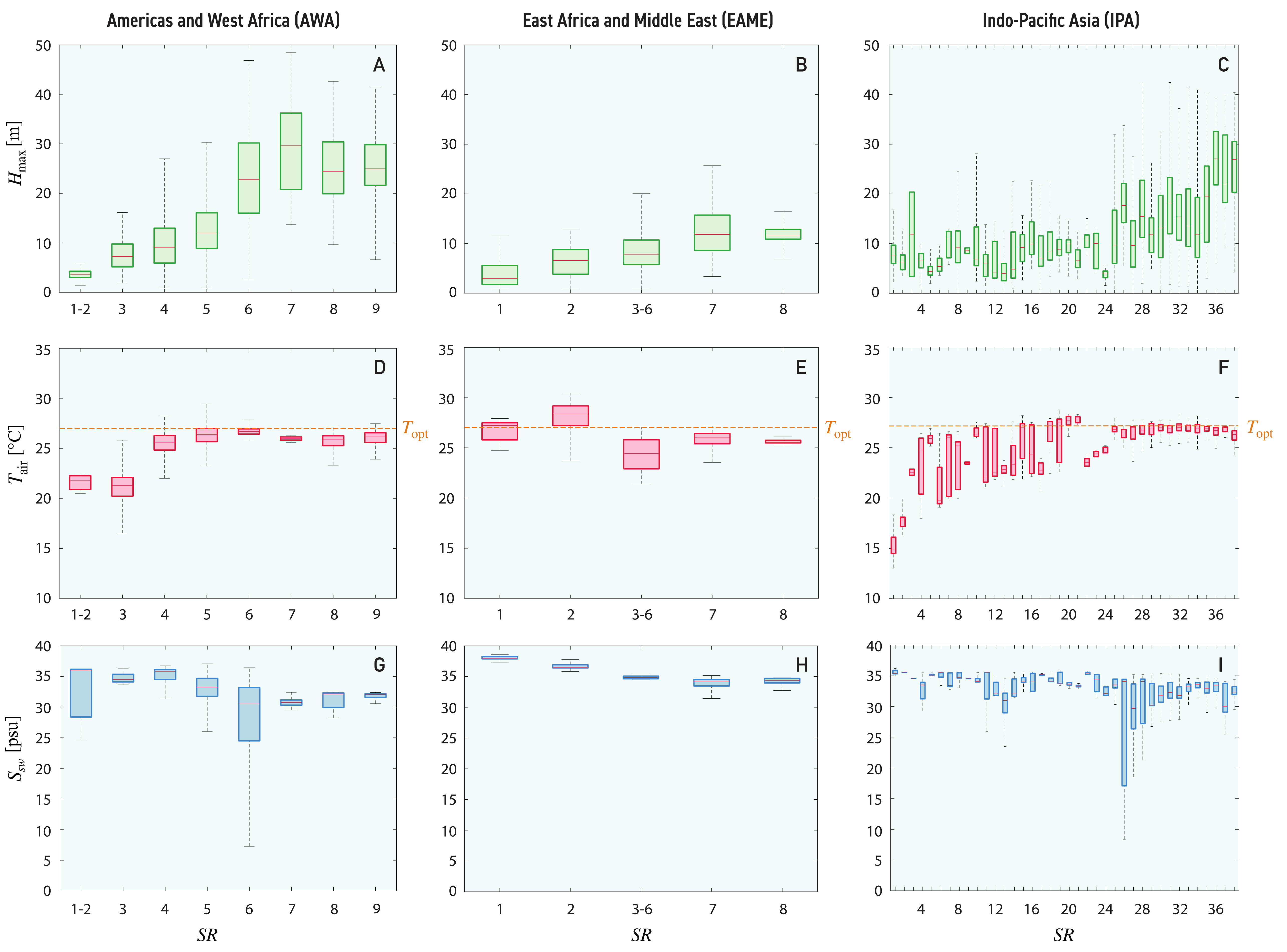}
 \caption{{\bf{Structure of the empirical relations between canopy height, air temperature, salinity, and species richness:}} Box plots of (A-C) maximum Mangroves canopy height [m], (D-E) salinity [psu], and (G-I) air temperature [$^{\circ}$C] or given species richness, $SR$, in the three considered sub-regions (AWA, EAME, and IPA). Red central marks indicate the median, and the bottom and top edges of the boxes indicate the 25$^{\rm{th}}$ and 75$^{\rm{th}}$ percentiles, respectively. The whiskers extend to twice the interquartile range, and the more extreme points are considered outliers (not shown). The optimal air temperature value corresponds to $T_{\rm{air}}=$27$^{\circ}$C circa.}
\label{fig:Figure3}
\end{figure*}

\indent To better investigate the role of biodiversity in regulating Mangrove canopy height, we explore the regional dependence of $H_{\rm{max}}$, $S_{\rm{sw}}$, and $T_{\rm{air}}$ from $SR$.
Consistently with results from terrestrial forests \citep{gatti2017exploring,Marks2016}, $H_{\rm{max}}$ shows a distinct dependence on species richness, sharply increasing with biodiversity (Figure \ref{fig:Figure3}A-C). Low diversity, in contrast, is associated with short canopies. Although the complexity of coastal ecosystems and the high number of confounding factors hinder our capability to disentangle some cause-effect relations, these findings seem to support the hypothesis that high species diversity can promote competition for light and resources, resulting in taller Mangrove trees \citep{macarthur1967theory,gatti2017exploring}.\\
%===================================================%
%     Figure 3 - species richness
%===================================================%
%======================================END FIGURE 3
\indent Additionally, in ecotones characterized by low salinity, tall forests create opportunities for short shade-tolerant species to coexist with tall and light-demanding Mangroves \citep{lugo1986Mangrove}.
It is also likely that the conditions that lead to high species richness promote high productivity and taller canopies \citep{lovelock2008soil,north2017cover}. Moderate salinity typical of intertidal estuarine ecotones, for example, may result in both high productivity and elevated species richness \citep{spalding2010world,Simard2019,Friess2019}. To better unravel the direct effect of salinity on canopy height from the indirect effect due to species richness, we also analyzed the average $H_{\rm{max}}$ for given $SR$ (see Section 3.3 and Figure \ref{fig:Figure4}).\\
\indent The analysis of the $T_{\rm{air}}$--$SR$ (Figure \ref{fig:Figure3}D-F) and $S_{\rm{sw}}$--$SR$ (Figure \ref{fig:Figure3}G-I) relations, at the same time, indicates that both $T_{\rm{air}}$ and $S_{\rm{sw}}$ are major abiotic co-factors in controlling species richness.
While elevated salt-stress and low temperature are generally coupled with low species richness, high diversity is attained in ecosystems with moderate $S_{\rm{sw}}$ and elevated $T_{\rm{air}}$, such as in the IPA Mangroves.
% \indent $SR$, in general, declines with increasing salinity. 
While air temperature is often the dominant constraint for species richness (especially in the IPA mangroves), $SR$ also tends to decline with increasing salinity. The effects of $S_{\rm{sw}}$ on $SR$ and $H_{\rm{max}}$ are particularly evident in the EAME region (Figure \ref{fig:Figure3}H), where salinity exert a major control on species diversity and canopy height while $T_{\rm{air}}$ is generally high~\citep{Almahasheer2016,Al-Yamani2017,adame2020Mangroves}.
The Mangroves growing along the coast of the Arabian Gulf and the Red Sea are a prominent example of such salinity impacts. Here, extreme salinity has led to mono-dominant forests, where only the most salt-tolerant -- and less productive -- species can survive. 
Once high-tolerant mono-dominant stands are established, Mangrove ecosystems reach a stable state, and salinity becomes a weaker driver of canopy height -- due to the intrinsic salinity resilience of these communities. 
Factors other than salinity thus become major drivers of the ecosystem's productivity.
This mechanism aligns with recent findings indicating nitrogen limitation -- and not salinity -- as the leading cause of Mangrove dwarfism on the  Central Red Sea coast~\citep{Anton2020}.\\
\indent Whether elevated temperatures represent an additional source of stress and an ecological constraint for EAME $H_{\rm{max}}$ \citep{Medina1999,Zahed2010}, salinity seems to be a dominant stress factor.
% The highest $SR$ is found in the IPA region \cite{Duke1998} and is associated with temperatures around an optimal value, $T_{\rm{opt}}$, corresponding to about 27$^{\circ}$C. 
Temperatures around an optimal value, $T_{\rm{opt}}$, corresponding to about 27$^{\circ}$C, in the IPA region harbour the highest richness~\citep{Duke1998}.
The emergence of this optimal value results from a) the reduction in light-saturated photosynthetic rates, stomatal conductance, and photosynthetic efficiency, known to occur in Mangroves at $T_{\rm{air}} <T_{\rm{opt}}$ \citep{Davis1940,Ball1988,saenger2002Mangrove,Kao2004,Barr2009}; and b) heat stress at $T_{\rm{air}}>T_{\rm{opt}}$ to which Mangroves adapt producing energetically-costly osmolytes such as proline \citep{liu2020proline}.
Ultimately, both extremely high and low temperatures reduce the biogeographical range of Mangroves, with very few species surviving in temperate or hot environments \citep{spalding2010world,Alongi2012}.
%%%%%%%%%%%%%%%%%%%%%%%%%%%%%%%%%%%%%
%               PLSR                %
%%%%%%%%%%%%%%%%%%%%%%%%%%%%%%%%%%%%%  
\subsection{Quantifying the impact of climatic and ecological forcing on $H_{\rm{max}}$}
\label{sec:Forcings}
\begin{figure*}[t!]
\centering 
\includegraphics[width=1.0\linewidth]{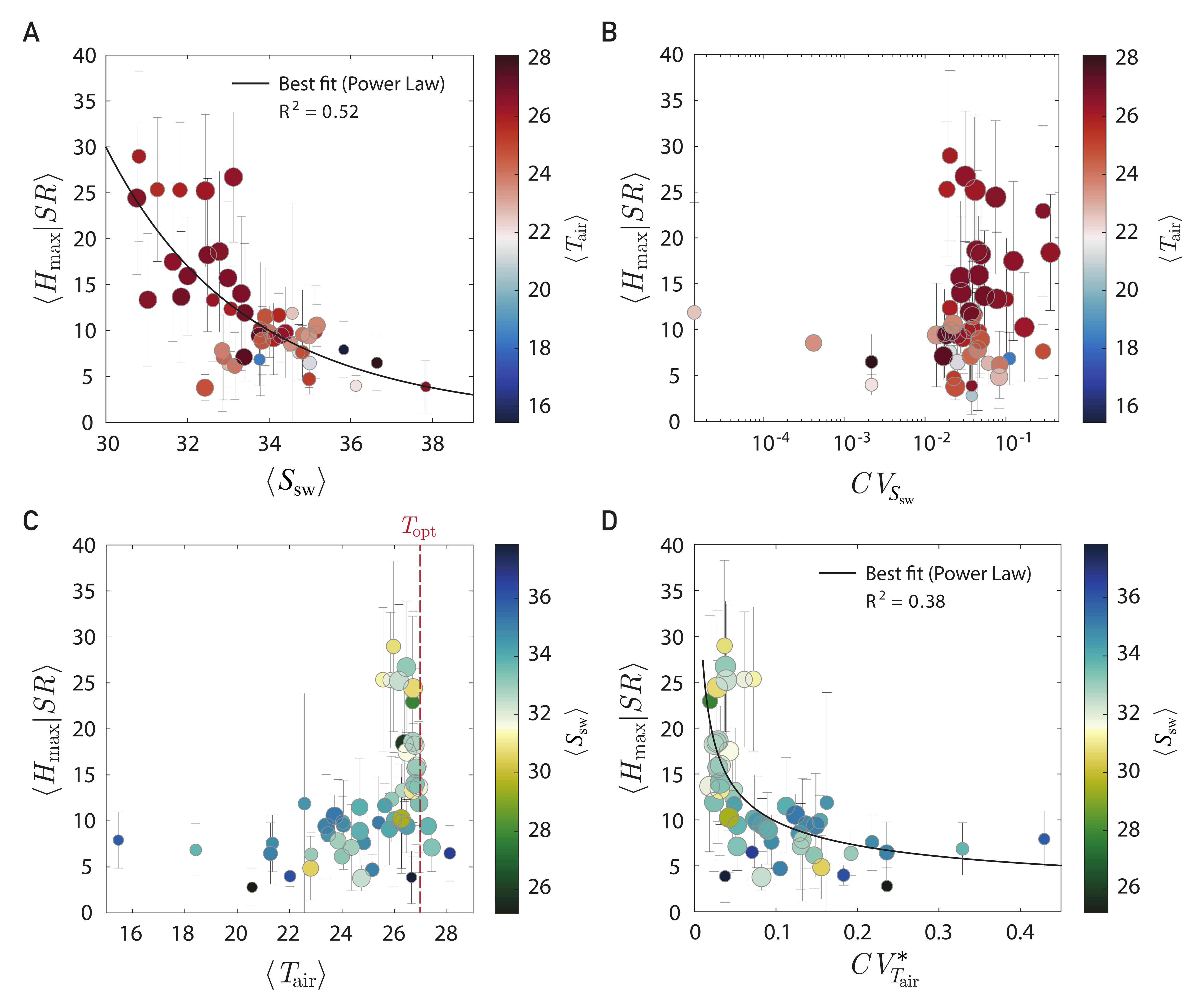}
\caption{{\bf{Average control of salinity and air temperature on Mangroves canopy height.}} a-b) Mean value of $H_{\rm{max}}$ for given $SR$, $\langle H_{\rm{max}} \mid SR \rangle$, as a function of a) mean $S_{\rm{sw}}$ and b) and salinity coefficient of variation, $CV_{S_{\rm{sw}}}$. The color bar refers to the mean value of $T_{\rm{air}}$. c-d) $\langle H_{\rm{max}} \mid SR \rangle$ as a function of c) mean $T_{\rm{air}}$ and d) and the modified temperature coefficient of variation relative to $T_{\rm{opt}}$, $CV^{*}_{T_{\rm{air}}}$. The color bar shows the mean value of $S_{\rm{sw}}$. Points size is proportional to $SR$, and the error bars represent the mean $\pm$ one standard deviation of the data.}
  \label{fig:Figure4}
\end{figure*}

\indent To assess the relative impact of $T_{\rm{air}}$, $S_{\rm{sw}}$, and $SR$ on Mangrove canopy height, the couplings observed across latitudes and at the local scales (Figures \ref{fig:Figure2}D-G and \ref{fig:Figure3}) were tested a) using a multivariate partial least-squares regression (PLSR; Tables \ref{tab:Table_1} and \ref{tab:Table_2}); and b) comparing mean $H_{\rm{max}}$ values for given $SR$, $\langle H_{\rm{max}} \mid SR \rangle$, to mean salinity and air temperature and their coefficient of variation (Figure  \ref{fig:Figure4}). 
$\langle H_{\rm{max}} \mid SR \rangle$ is computed as the average maximum canopy height across ecosystems of given species richness.
%===================================================%
%     Figure 4 - T and S controls
%===================================================%
%%%%%%%%%%%%%%%%%%%%%%%%%%%%%%%%%%%%%%%%%%%%%%%%%%%%%%%%%%%%%%%
The multivariate analysis estimates the percentage of $H_{\rm{max}}$ variance explained by the predictor variables, accounting for possible multicollinearity \citep{chong2005performance}. 
The analysis revealed that $SR$ is the main controlling factor for $H_{\rm{max}}$ in the latitude-dependent framework, explaining alone up to about 80$\%$ of the global variability (Table \ref{tab:Table_1}).
The percentage of variance explained by $SR$ at the regional level is significantly lower (41.8$\%$ -- 57.4$\%$) than the global one, suggesting that specific climatic factors become more important at the local scales. 
We find that, if partial correlations are taken into account, $T_{\rm{air}}$ is only weakly related to $H_{\rm{max}}$, and it does not significantly improve the performance of the regression model. These results suggest that the previously-reported impact of $T_{\rm{air}}$ on Mangrove canopy height \citep{Woodroffe1991,Duke1998,spalding2010world,Osland2017,Simard2019} may be attributed to the temperature-dependent geographical distribution of $SR$.\\
\indent Salinity is shown to be a crucial driver at the ecosystem scale controlling $H_{\rm{max}}$ both directly, as a source of environmental stress, and indirectly, by limiting biodiversity. 
Although $SR$ remains the most correlated variable also in the site-specific analysis explaining up to 34$\%$ of the variance in $H_{\rm{max}}$, $S_{\rm{sw}}$ becomes a significant predictor that explains up to 26$\%$ of the spatial variability (Table \ref{tab:Table_2}). 
The relation between $\langle H_{\rm{max}} \mid SR \rangle$ and $S_{\rm{sw}}$ shown in Figure \ref{fig:Figure4}A further supports the hypothesis that salinity can exert significant controls on canopy height. 
On average, Mangrove height decreases with salinity following a power law (R$^2$=0.52). 
In turn, the variability of $S_{\rm{sw}}$, quantified through its coefficient of variation $CV_{S_{\rm{sw}}}$ (see Methods), is not directly related to  $\langle H_{\rm{max}} \mid SR \rangle$ (Figure \ref{fig:Figure4}B), and  
displays a `threshold behavior' similar to the control shown by mean temperature (Figure \ref{fig:Figure4}C). 
Protracted and spatially homogeneous freshwater or hypersaline conditions result in low species richness \citep{Ball1998richness}. Yet, ecosystems with relatively low $SR$ (smaller dots in Figure \ref{fig:Figure4}A) can attain elevated heights if salt concentration is low, as a consequence of the coexistence of a moderate number of highly productive species. Variability in salinity and mean air temperature seem to impose ecological thresholds that determine biodiversity and only indirectly affect biomass stocks. 
This threshold can be easily identified for $T_{\rm{air}}$ around the optimal value $T_{\rm{opt}}$.\\
% %======================%
% %        Tables
% %======================%
% %%%%%%%%%%%%%%%%%%%%%%%%%%%%%% Table 1
% \begin{table}[t!]
% \caption{PLSR results, latitudinal analysis. The table displays the percentage of variance explained by the regression models at the global scale and for the different regions.}
% \begin{tabular}{  p{1.6cm}||p{2.0cm}|p{2.0cm}|p{2.0cm}|p{2.0cm}  }
%  \hline
% \textbf{Predictor} & Global & AWA & EAME & IPA\\
%  \hline 
% \textbf{$SR$ alone} & 80.4$\%$ & 57.4$\%$ &  41.8$\%$ &    51.7$\%$\\
% \textbf{$S_{\rm{sw}}$ alone} &  40.5$\%$ &   51.7$\%$  &  25.2$\%$   &  0.2$\%$\\
% \textbf{$T_{\rm{air}}$ alone} &  35$\%$ & 28.7 $\%$ &  7.1$\%$ &   28.6$\%$\\
% \textbf{Total}  &  85$\%$     & 74.6$\%$   & 42.8$\%$  & 61.4$\%$ \\
% $SR$,{\color{blue}$S_{\rm{sw}}$},{\color{red}$T_{\rm{air}}$} & (79.1,{\color{blue}4.6},{\color{red}1.2})    &  (62.4,{\color{blue}7.0},{\color{red}5.2})  &  (38.2,{\color{blue}4.3},{\color{red}0.3}) & (50.5,{\color{blue}2.2},{\color{red}8.6})\\
%  \hline
% \end{tabular} 
%  \label{tab:Table_1}
% \end{table}
\indent Interestingly, our results show that  $\langle H_{\rm{max}} \mid SR \rangle$ decreases as a function of the modified temperature coefficient of variation $CV^*_{T_{\rm{air}}}$, calculated as the ratio between the mean deviation around $T_{\rm{opt}}$ and the average air temperature (Figure \ref{fig:Figure4}D). Low $CV^*_{T_{\rm{air}}}$ thus indicates minimal temperature fluctuations around the optimal value. 
% The variability of $T_{\rm{air}}$ around $T_{\rm{opt}}$, represents another confounding factor that controls $H_{\rm{max}}$. 
This finding is in agreement with previous studies, including~\citet{Cavanaugh2019}, who observed an expansion of Mangrove ecosystems in the Florida Everglades as a consequence of the reduction in temperature fluctuations and freezing events frequency.\\
%======================%
%        Tables
%======================%
%%%%%%%%%%%%%%%%%%%%%%%%%%%%%% Table 1
\begin{table}[t!]
\caption{PLSR results, latitudinal analysis. All values are  averaged zonally over 1$^{\circ}$ of latitude across the globe, and for the AWA (120$^{\rm{o}}$ W - 13$^{\rm{o}}$ E), EAME (30$^{\rm{o}}$ E - 77$^{\rm{o}}$ E), and IPA (78$^{\rm{o}}$ E - 152$^{\rm{o}}$ W) region. The table displays the percentage of variance explained by the regression models at the global scale and for the different regions.}
\begin{tabular}{  p{1.6cm}||p{2.0cm}|p{2.0cm}|p{2.0cm}|p{2.0cm}  }
 \hline
\textbf{Predictor} & Global & AWA & EAME & IPA\\
 \hline 
\textbf{$SR$ alone} & 80.4$\%$ & 57.4$\%$ &  41.8$\%$ &    51.7$\%$\\
\textbf{$S_{\rm{sw}}$ alone} &  40.5$\%$ &   51.7$\%$  &  25.2$\%$   &  0.2$\%$\\
\textbf{$T_{\rm{air}}$ alone} &  35$\%$ & 28.7 $\%$ &  7.1$\%$ &   28.6$\%$\\
\textbf{Total}  &  85$\%$     & 74.6$\%$   & 42.8$\%$  & 61.4$\%$ \\
$SR$,{\color{blue}$S_{\rm{sw}}$},{\color{red}$T_{\rm{air}}$} & (79.1,{\color{blue}4.6},{\color{red}1.2})    &  (62.4,{\color{blue}7.0},{\color{red}5.2})  &  (38.2,{\color{blue}4.3},{\color{red}0.3}) & (50.5,{\color{blue}2.2},{\color{red}8.6})\\
 \hline
\end{tabular} 
 \label{tab:Table_1}
\end{table}
%%%%%%%%%%%%%%%%%%%%%%%%%%%%%% Table 2
\begin{table}[t!]
 \caption{PLSR results, site-specific analysis. All variables refer to the local (pixel) scale. Being highly correlated with species richness, $T_{\rm{air}}$ is not included in the multivariate analysis.}
\begin{tabular}{  p{1.8cm}||p{2.5cm}|p{2.5cm}|p{2.5cm}  }
 \hline
%  \multicolumn{4}{|c|}{Country List} \\
%  \hline
\textbf{Predictor} & AWA & EAME & IPA\\
 \hline
\textbf{$SR$ alone} & 33.8$\%$ &  30.4$\%$ &    16.4$\%$\\
\textbf{$S_{\rm{sw}}$ alone} &  26$\%$  &  6.7$\%$   &  0.35$\%$\\
\textbf{$T_{\rm{air}}$ alone} &  4.1 $\%$ &  7.1$\%$ &   4.4$\%$\\
\textbf{Total} &  59.8$\%$ & 37.1$\%$ & 16.8$\%$ \\
$SR$,{\color{blue}$S_{\rm{sw}}$} &  (33.8,{\color{blue}26.0})  & (30.4,{\color{blue}6.7}) & (16.4,{\color{blue}0.3})\\
 \hline
\end{tabular}
 \label{tab:Table_2}
\end{table}
%%==================================%%
%%           Conclusion             %%
%%==================================%%
\section{Conclusion}\label{Conclusion}
%%%%%%%%%%%%%%%%%%%%%%%%%%%%%%%%%%%%%%%%%%%%%%%%%%%%%
\indent We hypothesize that the salinity-induced limitation of canopy height results from a combination of different underlying processes, which can be summarized as follows. Analogously to water stress, salinity reduces soil water potential because of osmotic effects \citep{munns2002comparative,perri2017salinity,perri2018plant}. As canopy height increases, the osmotic stress intensifies the plant effort to sustain a leaf water potential that becomes more and more negative \citep{Koch2004,Klein2015}. As a result, salinity may impose a species-specific upper limit on canopy height, and Mangrove ecosystems experiencing prolonged and severe salt stress may tend to maintain a short canopy to avoid xylem cavitation \citep{Lovelock2006,Rossi2020}. \\ 
\indent Similarly, salinity limits carbon assimilation and biomass production \citep{munns2008mechanisms,Perri2019}. Salt-stressed plants prefer to allocate carbon in the root system at the expense of canopy expansion to maximize water use efficiency \citep{clough1982physiological,barr2013modeling}. Although the below-ground biomass of these ecosystems can become a significant carbon pool, above-ground carbon stocks and height are significantly limited \citep{Komiyama2008}.
We further argue that severe stress induced by salinity reduces the niche breadth by excluding less tolerant species and decreases competition for above-ground resources, limiting diversity and $H_{\rm{max}}$~\citep{Ball1998richness}. 
Contrarily, more tolerable salinity conditions might boost competition for light, enhance forest structural complexity and promote species coexistence, thus increasing canopy height and Mangrove productivity through complementary resource utilization \citep{Tilman2012,Grace2016}. 
The observed high $H_{\rm{max}}$ attained by tropical Mangrove ecosystems is, therefore, the outcome of the evolutionary competition for light and above-ground resources under favorable environmental conditions.\\
\indent Given the sharp gradients of salt concentration in coastal environments, salinity-induced limitations of $H_{\rm{max}}$ can have significant ecological implications under projected climate change and sea-level rise. 
A small increment in the sea level can lead to large sea-water intrusions \citep{Werner2009}, sizable soil salinisation \citep{fagherazzi2019sea}, and may reduce biodiversity and biomass stocks \citep{hayes2019groundwater,bai2021mangrove}. 
Failure in accounting for salinity impacts and salinisation-induced ecosystems shifts may lead to inaccurate predictions of Mangrove carbon stocks and species diversity, impairing our ability to design appropriate conservation measures to protect these vital ecosystems.

\section*{Acknowledgements}
This study was supported by the US Department of Energy Terrestrial Ecosystem Science Program under grant no. DE-SC0020116. S.P. and A.P. also acknowledge the support provided by the Princeton University’s Dean for Research, High Meadows Environmental Institute, Andlinger Center for Energy and the Environment, and the Office of the Provost International Fund.

\section*{Conflict of interest}
The authors declare that they have no competing financial interests.

\section*{Data availability}
\label{sec:DataAv}
% \href{http://www.fao.org/soils-portal/data-hub/soil-maps-and-databases/harmonized-world-soil-database-v12/en/}{FAO soil portal}.
% Mangrove canopy data were made available by the Oak Ridge National Data Archive (ORNL DAAC; https://doi.org/10.3334/ORNLDAAC/1665). 
Mangrove canopy data were made available by the Oak Ridge National Data Archive \href{https://doi.org/10.3334/ORNLDAAC/1665}{(ORNL DAAC).}
Sea surface salinity data in proximity of coastlines were obtained from the \href{http://marine.copernicus.eu}{GLORYS12V1 reanalysis product}, while the air temperature data were extracted from the \href{http://worldclim.org/version2}{WorldClim global dataset}.  
The global distribution of Mangrove species (60 species) was made available by the authors of the World Atlas of Mangroves \citep{spalding2010world}.

\section*{Author contributions}
S.P. and A.M. designed and performed research; S.P. analyzed data; M.D. and A.P. helped design the testing hypothesis and develop the statistical analysis; S.P. and A.M. wrote the paper with inputs from all the authors.

% \section*{Supporting Information}

% Supporting information is information that is not essential to the article, but provides greater depth and background. It is hosted online and appears without editing or typesetting. It may include tables, figures, videos, datasets, etc. More information can be found in the journal's author guidelines or at \url{http://www.wileyauthors.com/suppinfoFAQs}. Note: if data, scripts, or other artefacts used to generate the analyses presented in the paper are available via a publicly available data repository, authors should include a reference to the location of the material within their paper.

\printendnotes

% Submissions are not required to reflect the precise reference formatting of the journal (use of italics, bold etc.), however it is important that all key elements of each reference are included.
%\bibliography{sample}
%
%\begin{biography}
%
%\end{biography}

% \graphicalabstract{example-image-1x1}{Please check the journal's author guildines for whether a graphical abstract, key points, new findings, or other items are required for display in the Table of Contents.}

\end{document}